\documentclass[12pt]{iopart}

\usepackage{url}

\usepackage{graphicx}

\usepackage{amsfonts}
\usepackage{caption}
\usepackage{subcaption}
\usepackage[backend=bibtex, maxnames=12,date=year, url=false, doi=false, citestyle=numeric, eprint=true, isbn=false ]{biblatex}
\usepackage{csquotes}
\usepackage{xcolor}
\newcommand{\changes}[1]{{\color{black}#1}}

\usepackage{xspace}

\makeatletter
\DeclareRobustCommand\onedot{\futurelet\@let@token\@onedot}
\def\@onedot{\ifx\@let@token.\else.\null\fi\xspace}

\def\eg{\emph{e.g}\onedot} 
\def\ie{\emph{i.e}\onedot}

\def\etal{\emph{et al}\onedot}
\makeatother

\begin{document}

\title[VAEs with Structured Image Covariance  Applied to Compressed Sensing MRI]{VAEs with Structured Image Covariance Applied to Compressed Sensing MRI}
	\author{M A G Duff$^1$, I J A Simpson$^2$,    M J Ehrhardt$^1$  and N D F Campbell$^3$   		}
	\address{$^1$ Department of Mathematical Sciences, University of Bath, 
		Bath, BA2\ 7AY, UK}
   \address{$^2$ Predictive Analytics Lab, University of Sussex,  BN1\  9RH, UK}
   \address{$^3$ Department of Computer Science, University of Bath, 		Bath, BA2\ 7AY, UK}
\ead{e-mail: M.A.G.Duff@bath.ac.uk}
\vspace{10pt}
\begin{indented}
\item[]June 2023
\end{indented}
	%
	%



	\begin{abstract}
 
		\paragraph{Objective} This paper investigates how generative models, trained on ground-truth images, can be used \changes{as} priors for inverse problems, penalizing reconstructions far from images the generator can produce. The aim is that learned regularization will provide  complex data-driven priors to inverse problems while still retaining the control and insight of a variational regularization method. Moreover, unsupervised learning, without paired training data, allows the learned regularizer to remain flexible to changes in the forward problem such as noise level, sampling pattern or coil sensitivities in MRI.  
  \paragraph{Approach}  We utilize variational autoencoders (VAEs) that generate not only an image but also a covariance uncertainty matrix for each image. The covariance can model changing uncertainty dependencies caused by structure in the image, such as edges or objects, and provides a new distance metric from the manifold of learned images. 
  \paragraph{Main results} We evaluate these novel generative regularizers on retrospectively sub-sampled real-valued  MRI  measurements from the fastMRI dataset. We compare our proposed learned regularization against other \changes{unlearned regularization approaches and  unsupervised and supervised deep learning methods}.  
  \paragraph{Significance}  Our results show that the proposed method is competitive with other state-of-the-art methods and behaves consistently with changing sampling patterns and noise levels.      
		\end{abstract}

\vspace{2pc}

	\section{Introduction} \noindent
Compressed sensing Magnetic Resonance Imaging (MRI) provides the benefits of MRI with faster acquisition times, \changes{ providing accurate reconstructions from sparsely sampled k-space data}. Sophisticated mathematical reconstruction techniques allow for high-quality images to be produced with just a subset of the full MRI measurements, which are quicker to obtain. This reduces costs but can also improve image quality by reducing motion artifacts.    Mathematically, we seek to recover \changes{the MRI magnitude image,} $x\in \mathcal{X}\changes{=\mathbb{R}^d}$,   from observed measurements, $y\in \mathcal Y\changes{=\mathbb{C}^m}$. 
	The two are related by a linear forward model, $A:\mathcal{X}\rightarrow \mathcal{Y}$ 
	giving the equation
	$Ax=y$.    In MRI, $A$ is composed of a Fourier transform and a subsampling operator that takes just a subset of the Fourier data. This is an ill-posed inverse problem, where measurements are incomplete and so multiple solutions may exist.  
	
	This difficulty can be mitigated by 	incorporating prior information; we consider this to be given in the form of a regularizer, $\mathcal{R}$, in a variational regularization framework~\cite{Lustig2007sparseMRI, Knoll2011, Xu2021}
	\begin{align}
	\arg\min_{x} \, d(Ax,y) + \mathcal{R}(x) \label{Var-reg}
	\end{align}
	where $d:\mathcal{Y}\times\mathcal{Y}\rightarrow [0,\infty]$ is a similarity measure,  ensuring that the reconstructed image matches the data, and $\mathcal{R}:\mathcal{X}\to [0,\infty] $ is a regularization term which is small if the image satisfies some desired property. 
	
	 We extend recent work, including~\cite{Bora2017, Dhar2018, Habring2021, Tripathi2018, Duff2021} that consider the use of a generative machine learning model as part of the regularizer, called \textit{generative regularizers}. A \changes{latent} generative model is designed to take a sample from a distribution in a low-dimensional latent space and generate data similar to \changes{the} training distribution. Common deep generative models include variational autoencoders (VAEs)~\cite{Kingma2014}, generative adversarial networks (GANs)~\cite{Goodfellow2014} and normalizing flows~\cite{Rezende2015}. The generative model is trained on high-quality data, obtained from fully sampled measurements, with no knowledge of the forward model, $A$, or any sub-sampled data. The idea of generative regularizers is to penalize inverse problem solutions that are dissimilar to the learned distribution on ground truth, or high-quality reconstructed, images.

	The generative part of a VAE consists of a network, such that for any point $z \in \mathcal{Z}\changes{=\mathbb{R}^n}$, within the learned latent space, the network outputs a distribution on the image space $\mathcal{X}$,  \begin{align}
	p_{G,\Sigma}(x|z)= \mathcal{N}(x; G_{\theta}({z}), \Sigma_\theta(z) )  \label{structured noise}
	\end{align} for parameterised functions $G_{\theta}: \mathcal{Z}\rightarrow \mathcal{X}$ and $\Sigma_{\theta}: \mathcal{Z} \rightarrow \mathbb{R}^{d\times d}$, $d := \dim(\mathcal X)$.	  In a standard VAE model, $\Sigma_{\theta}$ is taken to be a multiple of the identity matrix, \ie~$\Sigma_\theta(z):= \rho I$, where the predicted variance, $\rho$, is either fixed or learned. This assumes that the reconstruction error of the generated image is independent and identically distributed for all pixels. In reality,  parts of an image, such as the background, will be easy to model whereas sharp edges \changes{and other high frequencies} are harder to model accurately. Similarly,  as images are commonly piece-wise smooth, there are often local correlations in the errors made by the model. A well-known issue with VAEs is their tendency to produce smooth images (see~\eg~\cite{Ruthotto2021}), missing the sharp edges that exist in real data.   In this work, we consider the effect of a more expressive covariance matrix, $\Sigma_{\theta}$, in a generative regularizer.

\paragraph{Contributions}
We propose \changes{an} adaptive generative regularizer where edge and correlations {in the data are modeled with a structured covariance network. We demonstrate the strength of this model by reconstructing knee MRI images from retrospectively sub-sampled real-valued data from the fastMRI dataset~\cite{Zbontar2018}.} \changes{Our contributions are outlined below.}

	\begin{itemize}
	 \item \changes{An adaptation of} the structured covariance model introduced by Dorta~\etal~\cite{Dorta2018}  for \changes{magnitude} images from the fastMRI dataset of   knee MR images.
	     \item An extension of the denoising example from Dorta~\etal~\cite{Dorta2018} for use in inverse problems with a non-trivial forward model, producing a novel generative regularizer.
	     \item A demonstration of how the prior provided by the VAE with structured covariance can be explicitly visualized. We see that the structured covariance model has learned long-range correlations between pixels, taking into account image structure.
	    \item An ablation study to compare three different options for the decoder covariance, $\Sigma_\theta(z)$: $\Sigma$ is a fixed constant multiplying the identity matrix, $\Sigma$ is a diagonal matrix with a learned diagonal and the proposed structured covariance where  $\Sigma$ is dense. We show the most flexible, dense covariance method, produces the best inverse problem results.
	    \item Comparisons of our proposed regularizer with a variety of regularizers:  the unlearned Total Variation (TV)~\cite{Rudin1992} and least squares \changes{with early stopping} methods. Also with \changes{three} unsupervised methods: a deep image prior approach with a pre-trained generator from Narnhofer~\etal~\cite{Narnhofer2019}, the original Compressed Sensing using Generative Models work by Bora~\etal~\cite{Bora2017} \changes{and a Plug-and-Play method by Ryu~\etal~\cite{ryu2019plug}}. Finally, with a state-of-the-art, learned, end-to-end method, variational networks~\cite{Hammernik2018}. We demonstrate that our method is competitive with the state-of-the-art methods, yet offers superior generalization to \changes{unseen} noise statistics and sampling patterns. 
	\end{itemize}

		\section{Related Work}
	Deep learning approaches to image reconstruction in inverse problems is a growing field~\cite{Arridge2019}. There are several supervised deep learning approaches~\cite{Zhu2018a, Hammernik2018, Oh2018,Wang2016,Quan2018a, Hyun2018, Mardani2019} that require datasets of subsampled measurements paired with high-quality reconstructions.   With any change in the forward model,~\eg a different k-space sampling pattern, or noise level,  new data needs to be acquired and models retrained. Furthermore, with these methods, care needs to be taken to ensure the image reconstruction is consistent with the observed measurements and these methods can also be unstable to small perturbations in the measured data~\cite{Antun2019}.
	In contrast, deep image priors~\cite{Ulyanov, Narnhofer2019}, have no data requirements and instead use an untrained convolutional neural network as a prior for the inverse problem. The prior is implicit and comes from the architecture choices, another choice to make, but also requires regularization in the form of early stopping to prevent over-fitting to the potentially noisy data.
	
	We choose to take an unsupervised approach, where we have example ground truth images but no paired data. The image modeling, for training the regulariser, and the forward modeling for the inverse problem reconstruction, are completely decoupled. We model the images using a generative model incorporating it as part of a regulariser for the inverse problem reconstruction. There has been previous work in this area~\cite{Bora2017, Dhar2018, Habring2021, Tripathi2018, Duff2021}, and we extend it with the addition of a structured image covariance network. \changes{We train our generative model without paired training data and without knowledge of the forward problem and point to the interesting works of \cite{zhang2021conditional} and \cite{Jalal2021a} for examples of these settings. Other unsupervised approaches include plug-and-play methods which iteratively call a learned denoiser within a larger optimization or inference algorithm~\cite{Ahmad2020, ryu2019plug}. }

	 When choosing and training a generative model for use in inverse problems, the generative model must be able to produce a whole range of possible solutions. A common issue with GANs~\cite{Goodfellow2014} is, that they do not capture the full distribution of images they were trained on.   This failure can be subtle~\cite{Arora2017} and therefore difficult to identify. GANs also do not have an encoder, and finding a latent space value that corresponds to a particular image is a  non-convex optimization problem with multiple local minima.  In comparison, a VAE is more suitable for use in generative regularizers because they can reconstruct images across the span of the training distribution although with the consequence of fewer high frequencies.  We consider VAEs over other generative models such as normalizing flows or invertible neural networks, both of which have been applied to inverse problems~\cite{Jalal2021a,Kingma2018}, because of the regularizing effect of a lower dimensional latent space in the VAE.  	Dorta~\etal~\cite{Dorta2018} proposed the use of a   structured covariance as part of a VAE for denoising, and the novel addition of this work is the application to MRI reconstruction.
  
  \changes{The Dorta~\etal~\cite{Dorta2018} approach parameterizes a dense covariance through the Cholesky decomposition of the corresponding precision matrix, the inverse covariance matrix, $\Sigma^{-1}_\theta$. Other approaches could include a diagonal plus low-rank parameterization of the covariance matrix, as seen for a segmentation task in \cite{NEURIPS2020_95f8d990}. While we believe a low-rank approximation may not give the fine detail required for image reconstruction, a combination of the two approaches could be an avenue for future work.  }

	\section{Method}
	
	 \noindent We build a probabilistic model for the reconstructed image, $x$, given an observation, $y$. First, \changes{we} consider the likelihood of the measurement, $y$, given image, $x$,  denoted $p(y|x)$, and usually taken to be $\mathcal{N}(y;Ax, \gamma^2I)$ for additive Gaussian noise over the observations with standard deviation, $\gamma$.  We choose a prior on the images $x$  given by a pre-trained generative model,  $p_{G,\Sigma}(x|z)$ as in~\eref{structured noise}. Finally, let $p_{\mathcal Z}$ be prior on the latent space used to train the generator, usually $\mathcal{N}(z; 0, I )$. 
	Combining these parts, we have that 
	\begin{align}
	p(x,z|y)&\propto p(y|x,z) \, p_{G,\Sigma}(x|z) \, p_{\mathcal Z}(z)\\ &=p(y|x) \, p_{G,\Sigma}(x|z) \, p_{\mathcal Z}(z)\label{heir}.
	\end{align}
\changes{Where the equality follows because $y$ is independent of $z$ given $x$. }	

	Ideally, we would seek to marginalize out the latent vector, $z$, however this integral is intractable, except by expensive sampling \changes{over the full distribution of $\mathcal{Z}$}; instead, we take a  maximum a posteriori (MAP) estimate.  By taking logarithms, maximizing~\eref{heir} with respect to $x$ and $z$  is equivalent to variational regularization~\eref{Var-reg} with
    \begin{align}
	 d(Ax,y)&=\frac{1}{2\gamma^2}\Vert Ax-y\Vert _2^2 \label{data}	\end{align} and 
	 \begin{align}
	 \mathcal{R}(x)&= \min_z \left(\log (|\Sigma_\theta(z)|)+ \frac{\|x-G_\theta(z)\|^2_{\Sigma_\theta(z)}}{2} +\frac{\Vert z\Vert _2^2}{2}\right), \label{reg}
	\end{align}
	where we denote the weighted norm by $\|x\|_M^2 := x^T M^{-1} x$ and the determinant of a matrix $\Sigma$ by $|\Sigma|$. Equation~\eref{data} 	 ensures that  $x$ explains the observation $y$, while the second term in~\eref{reg} constrains $x$ to be close to images in the range of the generator. 
	
	\paragraph{VAEs with Structured Covariance}
 The generative model is trained to minimize the distance between the generated, $p_{G,\Sigma}(\cdot; \theta)$, and training, $p_\text{Im}$, distribution. A VAE is derived by choosing the distance measure to be a  Kullback--Leibler divergence and then maximizing a lower bound approximation to this distance~\cite{Kingma2014}. The intractable  distribution over the latent space, $p_{G,\Sigma} ({z|x};\theta)$, is approximated by an encoder 
 \begin{align}
     q(z|{x};\psi)=\mathcal{N}(z; \mu_{\psi}({x}), \text{ diag}(\sigma^2_{\psi}({x}))) =:\changes{\mathcal{N}_{z, \psi}(x)}
 \end{align} with neural networks  $\mu_{\psi}, \sigma^2_{\psi}$, parameterized with weights, $\psi$. Following the derivation in~\cite{Kingma2014}, training a VAE becomes  a minimization with respect to $\psi$ and $\theta$ of  
	\begin{align}
    \mathbb{E}_{x\sim p_\text{Im}}\mathbb{E}_{z\sim \changes{\mathcal{N}_{z, \psi}(x)}} \left(  \text{nll}(x,G_\theta(z), \Sigma_\theta(z))   +\changes{d_{KL}(\mathcal{N}_{z, \psi}| p_{\mathcal Z}) }\right)\label{VAEs}
	\end{align} 
	where we write the negative log-likelihood as \begin{align}
	    \text{nll}(x,G ,\Sigma)=\log(|\Sigma|)+ \frac{1}{2}\|x-G\|^2_{\Sigma} + \text{constants} 
	\end{align} and the expectation over $x$ is calculated empirically over the training set \changes{via sampling}. The expectation over $z$ is also calculated empirically and for more details see the original paper~\cite{Kingma2014}.

	\paragraph{Structure of $\Sigma_\theta(z)$}

\begin{figure}
		\centering
		\includegraphics[width=.7\linewidth]{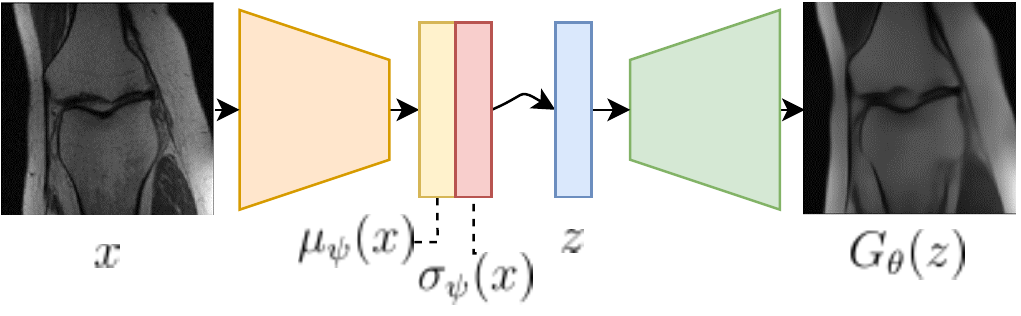}
		\includegraphics[width=.7\linewidth]{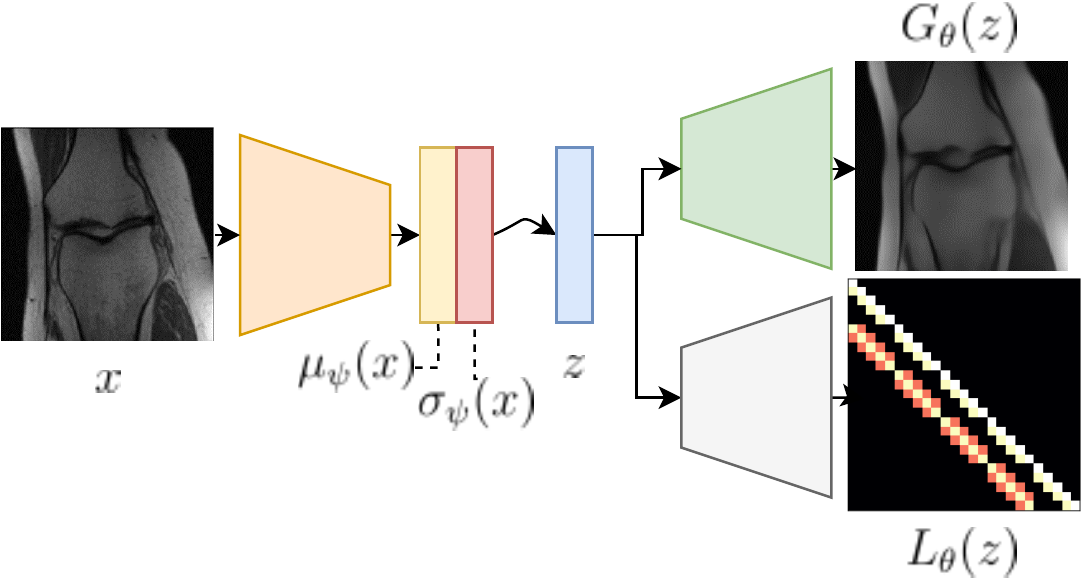}
		\caption{Comparison of the standard VAE (top) and the Structured Uncertainty Prediction Networks as developed by Dorta~\etal~\cite{Dorta2018} (bottom). The VAE has an encoder network that outputs a distribution, $\mathcal{N}(z; \mu_{\psi}({x}), \text{ diag}(\sigma^2_{\psi}({x})))$, which is then sampled from to get a latent vector, $z$, which is passed to the generator network, $G_\theta(z)$. In addition to the usual generator, the proposed network includes a network that takes latent vectors, $z$,  and outputs the weights of a sparse lower triangular matrix, $L_\theta(z)$. This matrix corresponds to the Cholesky decomposition of the inverse covariance for the generated image.}
		\label{fig:diagrams}
	\end{figure}

	Using a dense matrix $\Sigma_{\theta}$ is computationally infeasible for even moderate-sized images as it is expensive to calculate both the log determinant and the inverse required for the norm.  We use the computationally efficient Structured Uncertainty Prediction Networks as developed by Dorta~\etal~\cite{Dorta2018}. For each point in the latent space, an additional decoder, parameterized by $\theta$, outputs a sparse lower triangular matrix, $L_\theta(z)$, with the diagonal constrained to be positive. This forms the Cholesky decomposition of precision matrix, $\Sigma^{-1}_{\theta}$, such that  $\Sigma_{\theta} =(L_\theta L_\theta^T)^{-1}$. The Cholesky decomposition is taken to be sparse:     $(L_{\theta})_{i,j}=0$ if $j\neq N_i$ \changes{where $N_i$ is the set of neighbors of the pixel $i$}. This leads to a sparse precision matrix, $\Sigma^{-1}_{\theta}$. A zero value at entry $i$ and  $j$ in $\Sigma^{-1}_{\theta}$ means these pixels are independent, conditioned on all other image pixels. The pixels could still be correlated and thus this still allows for a dense covariance matrix. Indeed we would not choose to make $\Sigma_{\theta}$ sparse because zero values in the covariance indicate two uncorrelated pixels and we wish to allow  dependencies across images.  The local sparsity structure is also amenable to parallelization on the GPU and for more details see~\cite{Dorta2018a}. A pictorial view of the full network including an example sparse Cholesky matrix is given in figure~\ref{fig:diagrams}.

	\paragraph{Objective functions}
	This sparse Cholesky decomposition of the precision makes predicting a dense covariance matrix computationally feasible. The negative log-likelihood from~\eref{VAEs} becomes 
	\begin{align}
    \text{nll}(x,G, \Sigma)=-2\sum_{i=1}^d\log(L_{ii})+\frac{1}{2}\|L (x-G)\|^2_2 
	\end{align}
	up to constant terms and with, as above, $\Sigma^{-1}=LL^T$. The log determinant is now just a summation over the diagonals of the Cholesky matrix, we have removed the dense matrix inversion and there is no need to build the full covariance matrix. This also simplifies the regularizer from~\eref{reg}, giving 
	\begin{align}
	 \mathcal{R}(x)&=\min_z\left( \text{nll}(x,G_\theta(z), \Sigma_\theta(z)) +\frac{1}{2}\Vert z\Vert _2^2 \right). \label{regulariser}
\end{align}
	 Moreover, sampling from the extended VAE is possible by solving a sparse system of equations to get a sample $x=G_\theta(z)+(L_\theta(z)^T)^{-1}u$ where $u\sim \mathcal{N}(0,I)$.

	We note that the minimization problems in~\eref{VAEs} and~\eref{Var-reg} with~\eqref{regulariser} are not well defined as they are not bounded from below. Pixel values that can be determined with high accuracy~\eg those of a consistent black background, can have extremely low variance and the log determinant term may become large and negative.  \changes{We both bound the size of the diagonals of $L_\theta$ using a scaled tanh
activation and added a very small amount of noise to the black background to
deal with this.} 	Investigating other priors on  $L_\theta$ is an interesting area for future work.

	\section{Experiments \label{experiments}}
	\noindent	\paragraph{Dataset}
	The NYU fastMRI knee dataset contains, amongst other data, 796 fully sampled knee MRI magnitude volumes~\cite{Knoll2020a, Zbontar2018}, without fat suppression. This data was acquired on one of three clinical 3T systems
(Siemens Magnetom Skyra, Prisma and Biograph mMR) or one clinical 1.5T system, using a 15-channel knee coil array and conventional Cartesian 2D TSE protocol employed clinically at NYU School of Medicine. We use the ground truth data from the fastMRI single coil challenge, where the authors used emulated single-coil methodology to simulate single-coil data from the multi-coil
acquisition. {In addition, in the fastMRI dataset the images are cropped to a square, centering the knee.} We extract 3,872 training and 800 test ground truth \changes{slices} from this dataset, selecting images from near the center of the knee,  resize the images, using the Python imaging library~\cite{Clark2015} function with an anti-aliasing filter,  to $128\times128$ pixels and linearly rescale each image to the range $[0,1]$. The training and test datasets correspond to the training and validation sets of the fastMRI dataset and images from the same volume are always contained in the same dataset.

	\paragraph{Forward Problem}
	Our forward problem is inspired by the fastMRI single-coil reconstruction task; reconstructing images approximating the ground truth from under-sampled single-coil MR data~\cite{Zbontar2018}. The  ground truth images are Fourier transformed and subsampled. {To this end, a mask selects the points in k-space corresponding to a sampling pattern. We use both radial and cartesian sampling patterns, selecting radial spokes and horizontal lines in k-space, respectively. }  	We use the operator discretization library (ODL)~\cite{Adler2017} in Python and take the same radial subsampling MRI operators as in~\cite{Bungert2020b}. {Note that this is a relatively simple MRI model, yet a good starting point to test the feasibility of the proposed approach. We discuss its limitations in section~\ref{discussion}}.

	\paragraph{Model Architecture}
	See figure~\ref{fig:diagrams} for a comparison of the VAE and the VAE with structured image covariance.  All the networks are built of resnet-style blocks\changes{,} which consist of a convolutional layer with stride 1, a resizing layer, followed by two more convolutional layers, and then a relu activation function. The output of this process is then added to a resized version of the original input to the block.   The resizing layer is either a bilinear interpolation for an up-sampling layer, increasing width and height by a factor of 2;  convolutions with stride 2 for a down-sampling layer or a convolution with stride 1 for a resnet block that maintains image size. We choose a latent space of size 100. The generative network consists of a single dense layer outputting 8x8 images with 16 channels before a resnet block without resizing to give 8x8 images with 256 channels, then four up-sampling blocks are applied giving image sizes 16x16, 32x32, 64x64 and 128x128 with channels 512, 256, 128 and 64  and final another resnet block without resizing to reduce the channels down to one output image.   The covariance network is identical but outputs a 128x128 image with 5 channels, from which the sparse Cholesky matrix is formulated, based on code from Dorta~\etal~\cite{Dorta2018}\footnote{\url{https://github.com/Era-Dorta/tf_mvg}}.  The sparsity pattern is chosen such that the Cholesky matrix $L_\theta(z)_{i,j}$ is non-zero only if pixels $i$ and $j$ lie in the same $5\times 5$ patch.   The encoder is reversed copy of the generator, with down-sampling layers replacing the upsampling layers.   We include drop-out layers during training.

		\paragraph{Model Training}
	Training the VAE with structured covariance is done in a two-stage process. First the generated mean, $G_\theta$, and the encoder, $\changes{\mathcal{N}_{z, \psi}(x)}$, are trained with a standard VAE loss~\cite{Kingma2014} \ie $\Sigma_{\theta} = \rho I$ where $\rho$ \changes{is a fixed hyperparameter}.  The weights for the mean \changes{decoder} and \changes{the} encoder are then fixed before the covariance model is trained using \changes{the full structured noise loss given by}~\eref{VAEs}. The choice of this two-stage training is two-fold: firstly, it forces as much information as possible to be stored in the weights of the mean, and not the covariances, and secondly, it allows us to compare the effect of just changing the covariance models in the ablation study.  	Models were built and trained in TensorFlow using an NVIDIA RTX 2080 - 8GB GPU. It took approximately 8 hours to train the means and then 24 hours and 30 hours for the diagonal and structured covariance models.  The code will be available on publication.

		\paragraph{Ablation Study} We consider three variations on the structure of $\Sigma_\theta(z)$: $\Sigma_\theta$ is a fixed constant multiplying the identity matrix, $\Sigma_\theta$ is a diagonal matrix with a learned diagonal and our proposed method where $\Sigma_\theta$ is a dense matrix.  We call these options: \textit{mean+identity}, \textit{mean+diagonal} and \textit{mean+covar*}. The \textit{mean+identity} model is just taken to be the output of the first part of the \textit{mean+covar*} training described above. For \textit{mean+diagonal}, we again take the learned generated mean, $G_\theta$, and the encoder, $\changes{\mathcal{N}_{z, \psi}(x)}$ from the \textit{mean+identity} model and then optimize~\eref{VAEs} for the covariance network, but choose the off diagonals of the Cholesky matrix, $L_\theta(z)$, to be zero, so that the final covariance matrix is diagonal.

	\paragraph{Proposed reconstruction method}
	For the proposed method \textit{mean+covar*} and the versions \textit{mean+identity} and \textit{mean+diagonal} we choose a variation on the above regulariser~\eref{regulariser}
	\begin{align}
	 \mathcal{R}(x)&=\min_z \lambda\left( \text{nll}(x,G_\theta(z), \Sigma_\theta(z)) +\frac{\mu}{2}\Vert z\Vert _2^2 \right). \label{reg2} 
	\end{align} 
	The addition of the two regularization parameters $\lambda$ and $\mu$ is in recognition that \changes{the optimization problem in \eref{regulariser}} is a non-convex problem. 
	We use alternating gradient descent with backtracking line search (see~\eg Algorithm 9.2 of~\cite{Boyd2004})  where gradient descent steps are taken, alternating in the $x$ and $z$ space, with step size chosen to insure the objective decreases. 	We initialize at a rough reconstruction, given by the adjoint of the forward operator,  and the encoding of the adjoint for $x$ and $z$, respectively. Regularization parameters were selected via a grid search to maximize the peak signal--to--noise ratio (PSNR).
	
	\paragraph{Unlearned method comparisons} We  compare against Total Variation (\textit{TV}) regularization~\cite{Rudin1992} implemented using the Primal-Dual Hybrid Gradient method~\cite{Chambolle2011}, with regularization parameter chosen to maximize PSNR. As a baseline, we also calculate the least squares solution, $\min_x\|Ax-y\|_2^2$, optimized using gradient descent with backtracking line search.  
	
	\paragraph{Data driven prior comparisons} For another example of a generative regulariser, we take our trained mean generator $G_\theta$ and implement the method of Bora~\etal~\cite{Bora2017},  minimizing with respect to $z$ the objective  \begin{align}
	    L(z)=\frac{1}{2}\Vert AG(z)-y\Vert _2^2 +\mu\Vert z\Vert _2^2 \label{Bora},
	\end{align} searching for an image in the \textit{range} of the generator that best matches the measurements. The regularization parameter, $\mu$, is chosen to maximize PSNR.  
	We also implement the approach of~\cite{Narnhofer2019} which takes a trained generator but then,  after observing data, tweaks the weights of the network, optimizing
	$z^*,\theta^* \in \arg\min ||AG_\theta(z)-y||_2^2$, we call this \textit{Narnhofer19}. We use the same mean generator as in our other experiments,  gradient descent with backtracking for the $z$ optimization and TensorFlow's inbuilt Adam optimizer for the network weight update. We choose 1000 iterations to find an initial value of $z$ and then select the number of iterations for the alternating $z$ and $\theta$ updates via a grid search to maximize the PSNR. 

 \changes{Finally, we take the method of Ryu~\etal \cite{ryu2019plug}, specifically their   Plug-and-play alternating direction method of multipliers (PnP-ADMM) algorithm, which takes an iterative ADMM approach but replaces the proximal function in the image space with a learned denoiser. Specifically, following the derivation in \cite{Boyd2011} we have
 \begin{align}
	x_k&\in \arg\min_x\left\lbrace\frac{1}{2\sigma^2}\mathcal{D}(Ax,y)+\frac{1}{2\eta}\|x-u_{k-1}+v_{k-1}\|\right\rbrace,\label{PnP-ADMM-1}\\ 
	u_k&\in\arg\min_x\left\lbrace \frac{1}{2\eta}\|x_k-v_{k-1}-x\|_2^2+\mathcal{R}(x)\right\rbrace, \label{PnP-ADMM-2}\\
	v_k&=v_{k-1}+(x_k-u_k).\label{PnP-ADMM-3}
\end{align}
The first equation \eref{PnP-ADMM-1} can be solved analytically and, in \eref{PnP-ADMM-2}, instead of specifying a regularisation term, the whole term is replaced by a learned denoiser. The variables $\sigma$ and $\eta$ are constants, $x_0$ is  a rough reconstruction given by the adjoint operator and  $u_0$ and $v_0$ are initialised at zero.
 We use the code implemented by the authors\footnote{\url{https://github.com/uclaopt/Provable_Plug_and_Play}}  for both training the denoiser and for the iterations. We take their RealSN-DnCNN denoiser using the default settings, training on the fastMRI dataset as with the other methods.  When running the iterative method on measured data, we choose the number of iterations to maximize PSNR. 
	}
	\paragraph{Supervised end-to-end method comparison} Finally, we also demonstrate comparisons with a  variational network (VN)~\cite{Hammernik2018}. This is a end-to-end approach  which takes~\eref{reg} with $d(Ax,y)=\frac{\lambda}{2}\|Ax-y\|^2_2$ and treats the regulariser $\mathcal{R}$ as some unknown function. Optimizing~\eref{reg} by gradient descent  will lead to updates of the form 
\begin{equation}
x_{t+1}=x_t-\alpha_t\lambda A^T(Ax_t-y)+\nabla\mathcal{R}(x_t),
\end{equation} where $A^T$ is the adjoint of $A$ and $\alpha_t$ is a step-size.  The authors replace the unknown $\nabla\mathcal{R}(x_t)$ term with a learned component, inspired by a Fields of Experts model~\cite{Roth2009}. Fixing the number of iterations and unrolling leads to an end-to-end method that includes information from the forward and its adjoint. 
 We use the same network design and parameters as the original paper. The iterations are first initialized with a rough reconstruction given by the adjoint. The gradient of the regularizer consists of a convolution with 24 filters, kernel size of 11 and stride 1; a component-wise activation function; and then a convolution transpose with the same dimensions to return an image with the same dimensions as the input. The activation function consists of a weighted sum of  31 radial basis functions with learnable means, variance and weights.  All learnable parameters are allowed to vary independently in each layer and the step-size $\alpha_t$ and parameter $\lambda$ are also learned.    We train three variations on our data: 25 radial spokes with 0.05 and 0.0125 added noise and 125 radial spokes with 0.05 added noise.  
	
	
	\section{Results}
\begin{figure*}
		\centering
		\includegraphics[width=.75\linewidth]{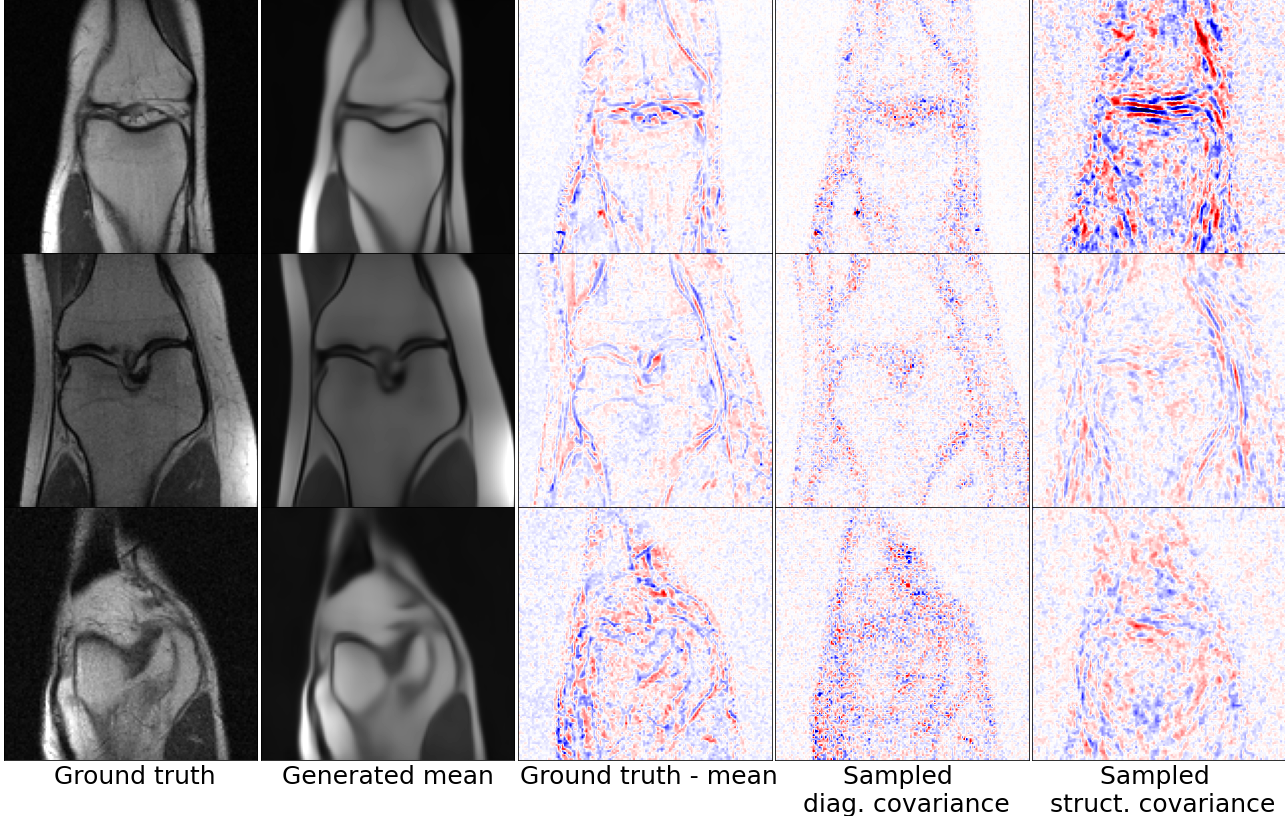}
		\caption{Comparing the covariance models \textit{mean+diagonal} and \textit{mean+covar*}. The first column gives the ground truth, and the second column a reconstruction in the \textit{range} of the VAE generator. The third gives the residual which can be considered as one sample of a zero mean Gaussian distribution with unknown covariance. The final two columns give  single residual samples from our learned covariances  \textit{mean+diagonal} and \textit{mean+covar*} models \changes{calculated using $(L_\theta(z)^T)^{-1}u$ where $u$ is one sample from $\mathcal{N}(0,I)$ and $L_\theta(z)$ is diagonal and lower triangular for the \textit{mean+diagonal} and \textit{mean+covar*} models, respectively}.   Columns are rescaled for better visualization.}
		\label{fig:plot_generative_models}
	\end{figure*}

	\noindent 	\paragraph{Ablation Study}
	To compare the covariance models \textit{mean+identity}, \textit{mean+diagonal} and \textit{mean+covar*}, visualizations of samples from the learned covariances are given in figure~\ref{fig:plot_generative_models}.   We see that the \textit{mean+diagonal} places uncertainty in the right locations,  but without structure, the samples are speckled and grainy. The \textit{mean+covar*} models can match some of the missing structures from the generated mean images. We note that these are random samples and therefore not expected to match the residual precisely but rather illustrate statistical similarity.

\begin{figure*}
		\centering
		\includegraphics[width=0.45\linewidth]{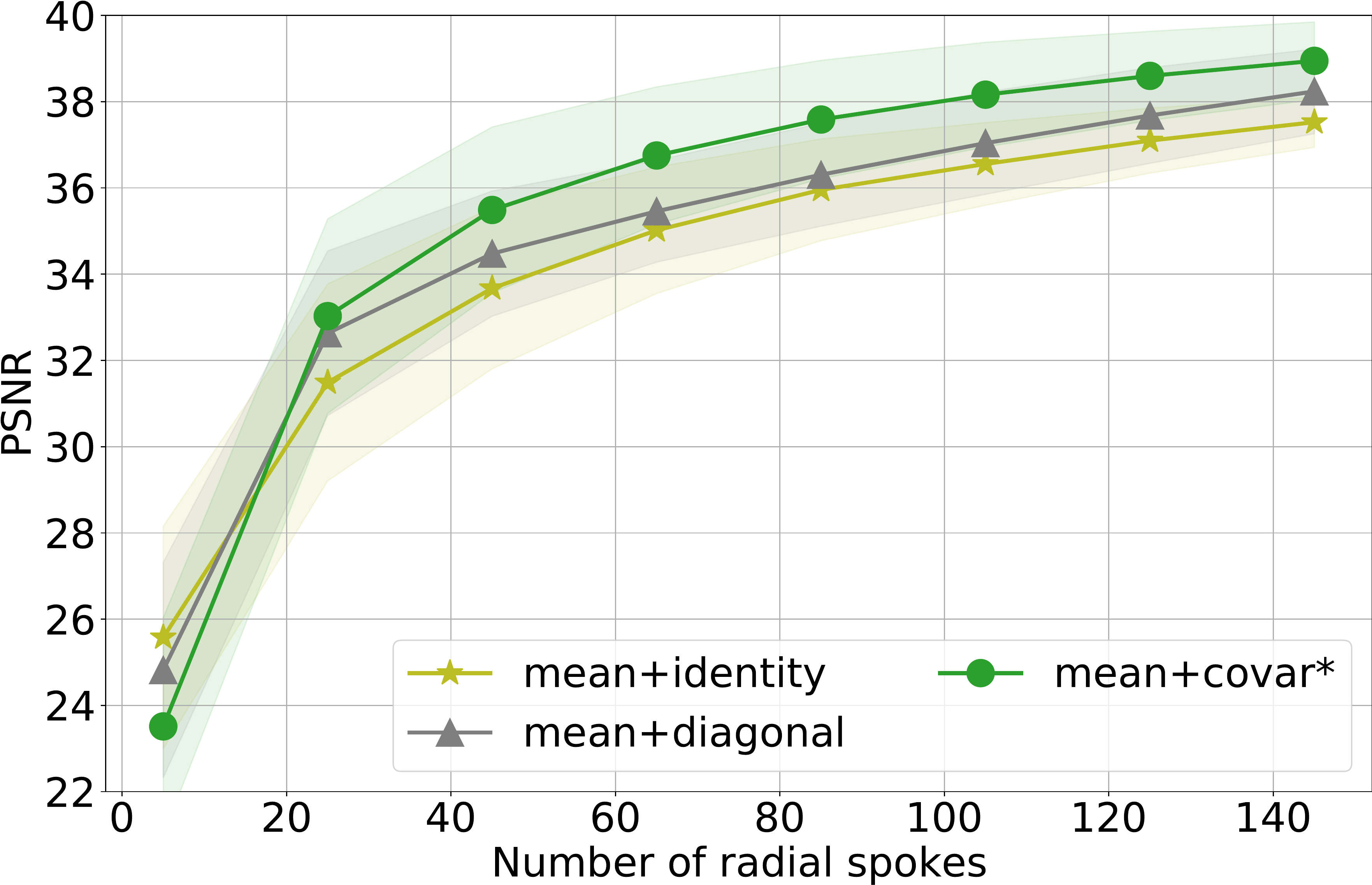}
		\includegraphics[width=0.45\linewidth]{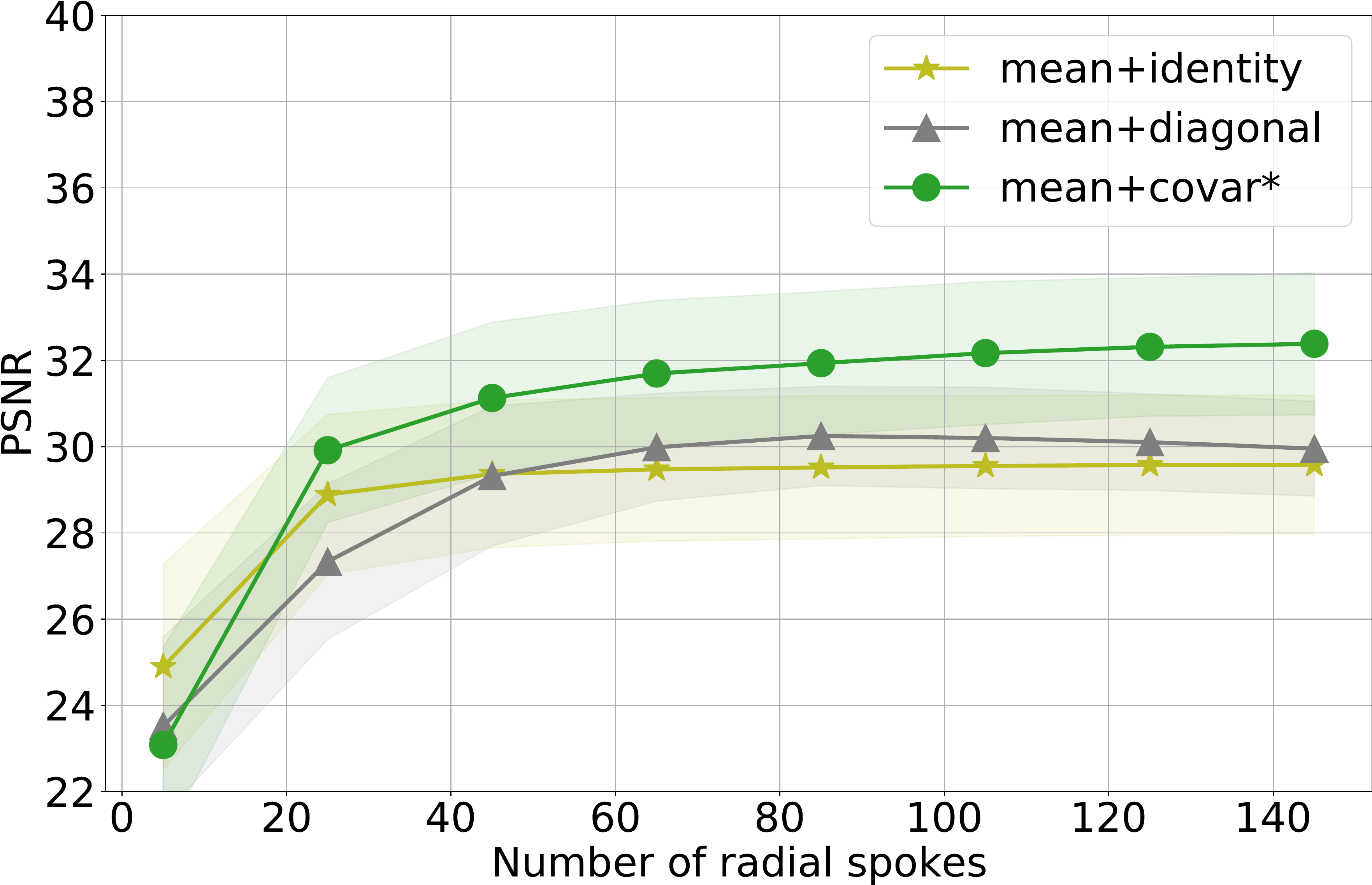}
		\caption{ Ablation study comparing generative regularisers based on VAEs with different noise models \textit{mean+identity}, \textit{mean+diagonal} and \textit{mean+covar*}. The plots show PSNR values average over reconstructions of 50 test images. The mean is given by the solid line and standard deviation in the shaded box.  The x-axis measures the number of radial spokes in the sampling pattern and the measured data was corrupted with additive Gaussian noise of standard deviation 0.0125 on the left and 0.05 on the right. }
		\label{fig:ablation}
	\end{figure*}

	Figure~\ref{fig:ablation}  compares the effectiveness of the three covariance models as generative regularisers.  The same observed data was used for each reconstruction method for each image. 
   The results for \textit{mean+covar*} give a consistently higher PSNR value than \textit{mean+diag} and \textit{mean+identity} and we use this method throughout the rest of the results section. These results support our hypothesis that learning a more flexible distance measure allows us to better fit the data and gives a better regularizer for our inverse problem. 

\begin{figure}
		\centering
		\includegraphics[width=.3\linewidth]{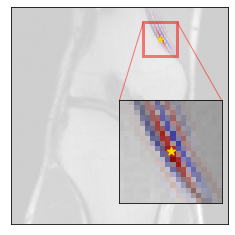}
		\includegraphics[width=.3\linewidth]{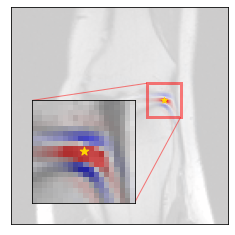}
		\caption{	Visualization of learned covariances between example pixels (yellow stars) and other pixel locations for the \textit{mean+covar} model. Red indicates a high positive correlation, and blue is a strong negative correlation.}
		\label{fig:learned-covar}
	\end{figure}

\paragraph{Prior introspection}
	We can explicitly examine the learned structured covariance, to visually assess the prior information passed to the generative model. To do this, we take a test image, $x$, and use the encoder to give a latent vector, $z$, that corresponds to the test image. From this we can calculate the structured covariance, $\Sigma_\theta(z) \in \mathbb{R}^{d\times d}$, where $d=\dim{\mathcal{X}}$. Each row of the covariance matrix corresponds to a pixel in the generated image, and the row can be reshaped into an image, showing the correlations between the chosen pixel and all others. Two example images with chosen pixels highlighted with a star are shown in figure~\ref{fig:learned-covar} \changes{and further examples are given in a supplementary video\footnote{\url{https://www.youtube.com/watch?v=_bi2D7rJ0OA}}}. Positive correlations are given in red, and negative correlations are in blue. We can see that the structure of the edges is present, and that, despite the local structure of the precision matrix, longer-range correlations have been learned. 

	\paragraph{Comparison with unlearned methods} 
	
		\begin{figure*}
		\centering
		\includegraphics[width=\linewidth]{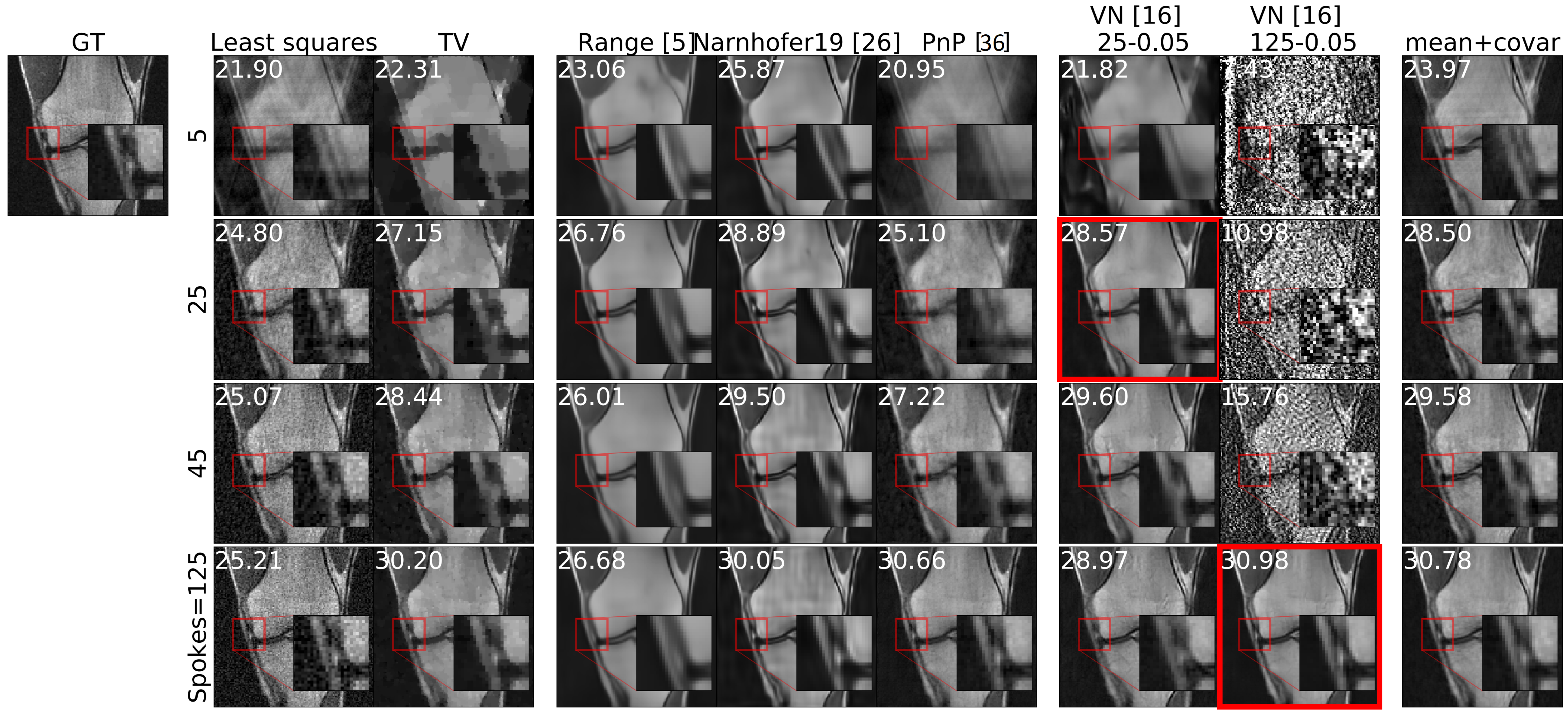}
		\caption{Example reconstructions of a test image for different amounts of measured data:  5, 25, 45 and 125 radial spokes, all with additive Gaussian noise with standard deviation 0.05. The columns give different reconstruction methods. The PSNR values are added in white and the red boxes indicate the settings the highlighted variational network has been trained on.}
		\label{fig:Images example}
	\end{figure*}

	\begin{figure*}
		\centering
		\includegraphics[width=\linewidth]{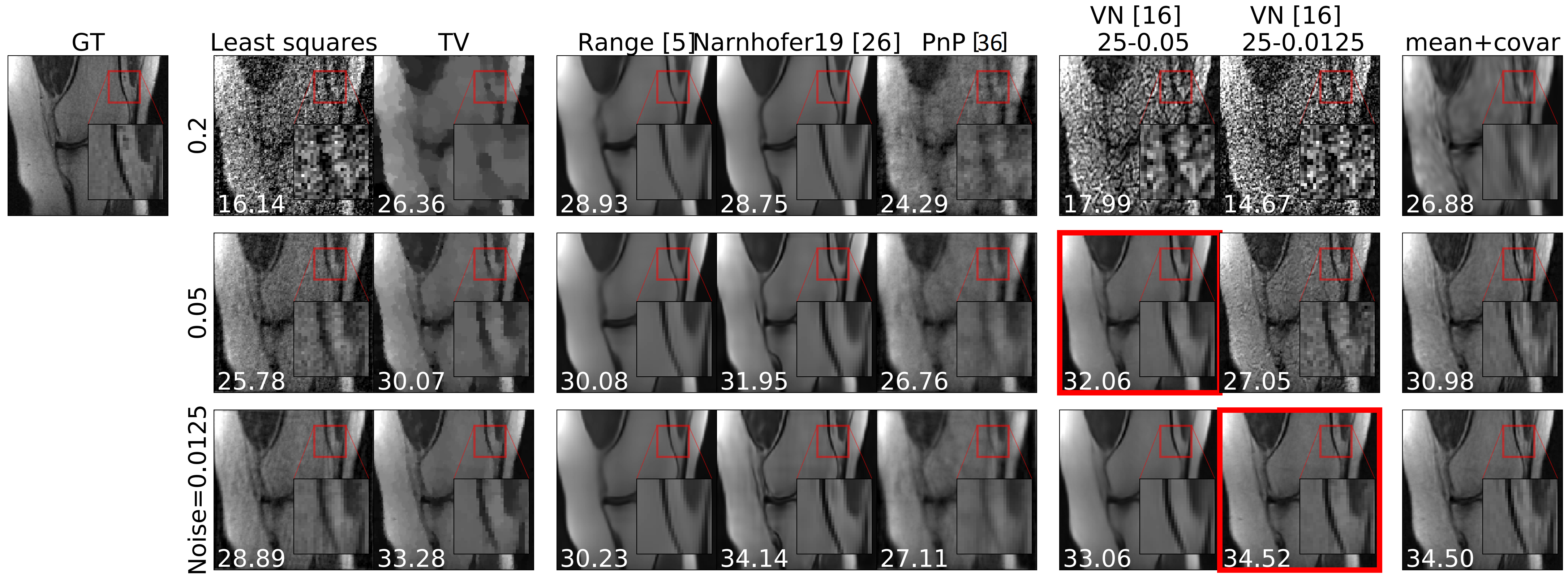}
		\caption{Example reconstructions of a test image with additive noise of different standard deviations:  0.2, 0.05 and 0.0125, all with a sampling pattern with 25 radial spokes.  The layout is as in figure~\ref{fig:Images example}.}
		\label{fig:Images example noise}
	\end{figure*}

\begin{figure*}
		\centering
		\includegraphics[width=0.45\linewidth]{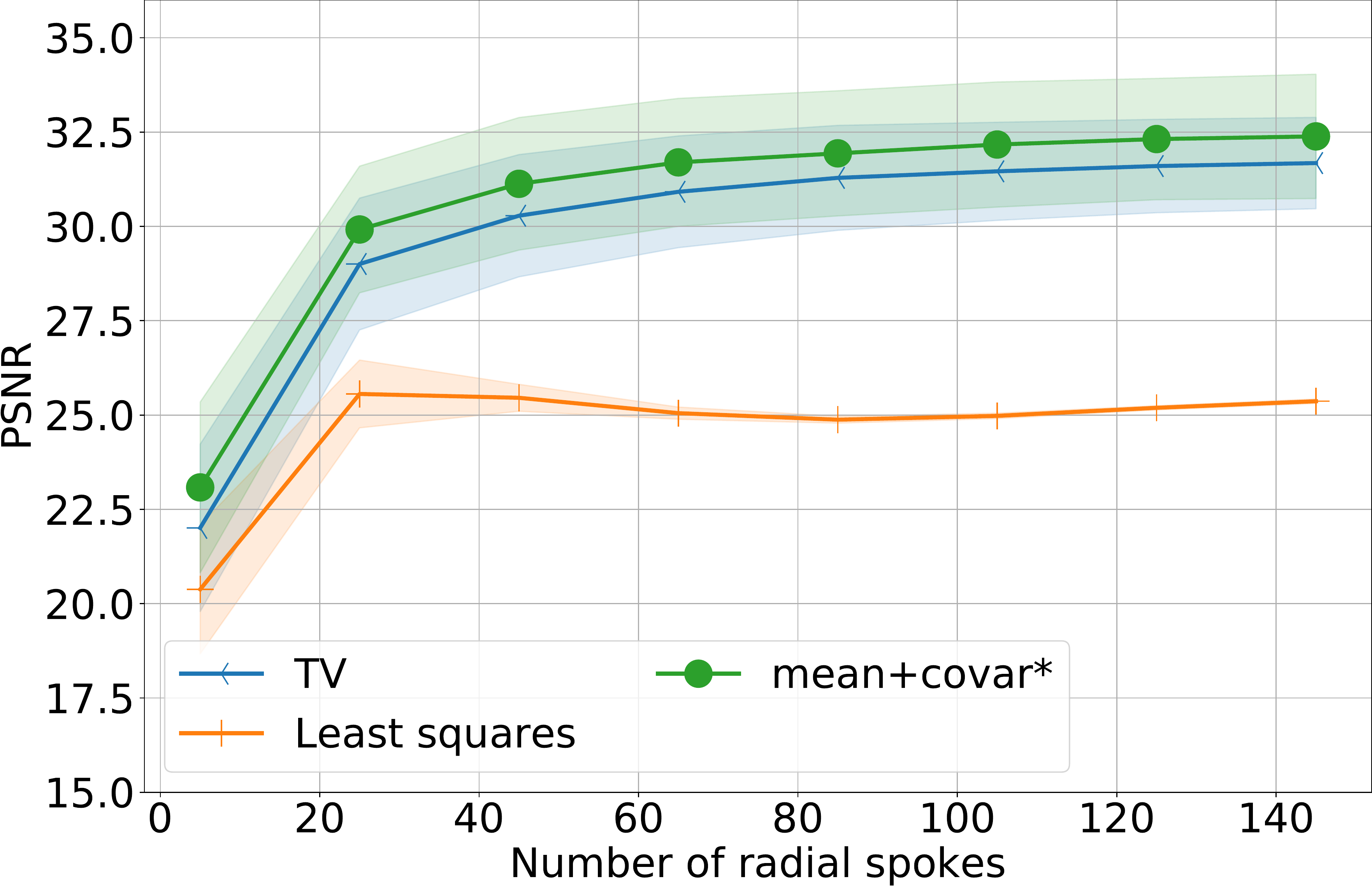}
		\includegraphics[width=0.45\linewidth]{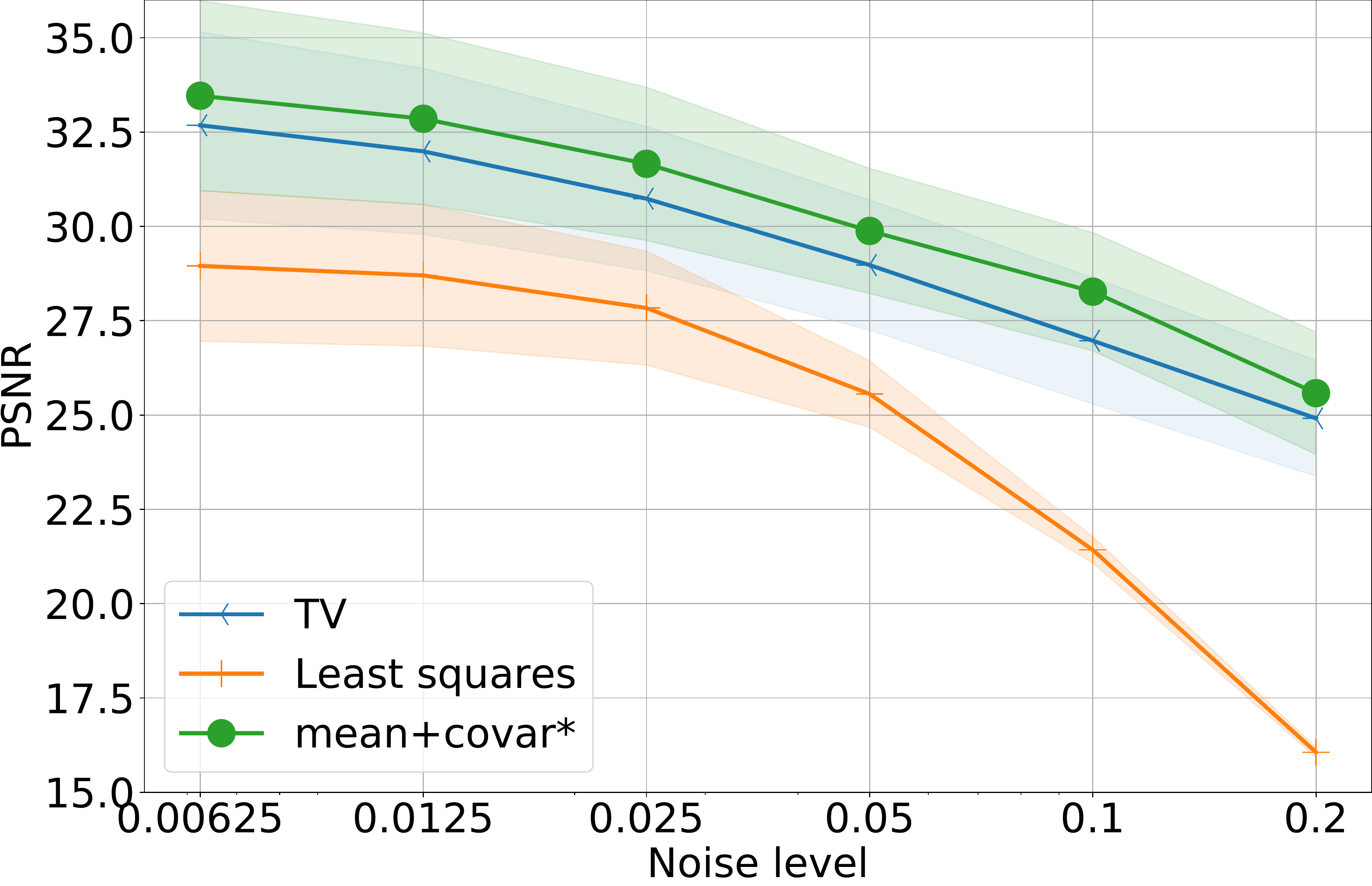}
		\caption{Comparison of the \textit{mean+covar*} method with the training-free methods TV and least squares.  The plots show the mean PSNR values for reconstructions averaged over 50 test images, with the standard deviations given by the shaded area. The left plot shows results from measured data with additive Gaussian noise of standard deviation 0.05 and differing numbers of measured radial spokes. The right plot gives results where 25 radial spokes are used for each reconstruction but the standard deviation of the added noise varies.   }
		\label{fig:comparepsnr-unlearned}
	\end{figure*}

\begin{figure*}
		\centering
		\includegraphics[width=0.45\linewidth]{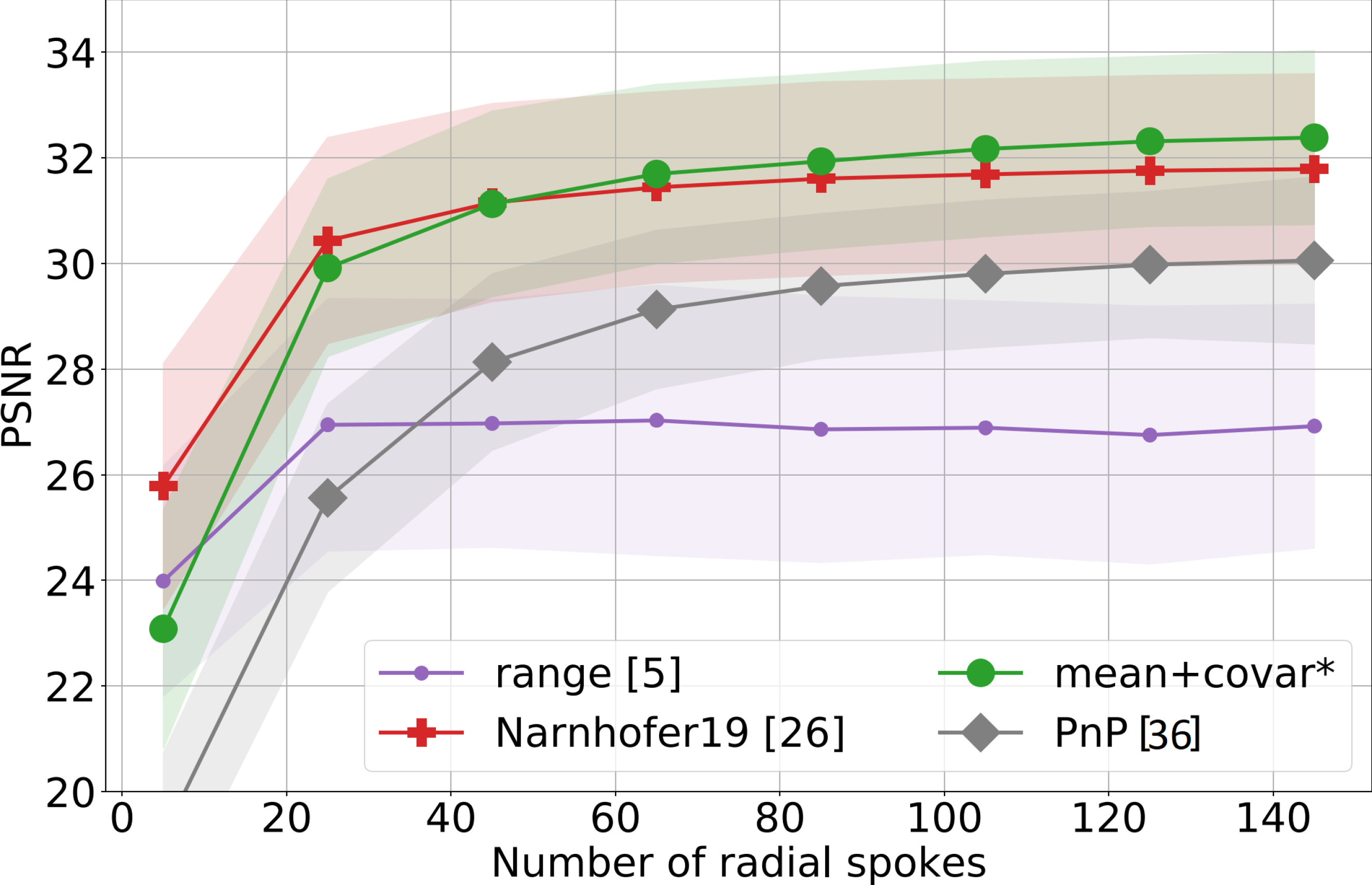}
		\includegraphics[width=0.45\linewidth]{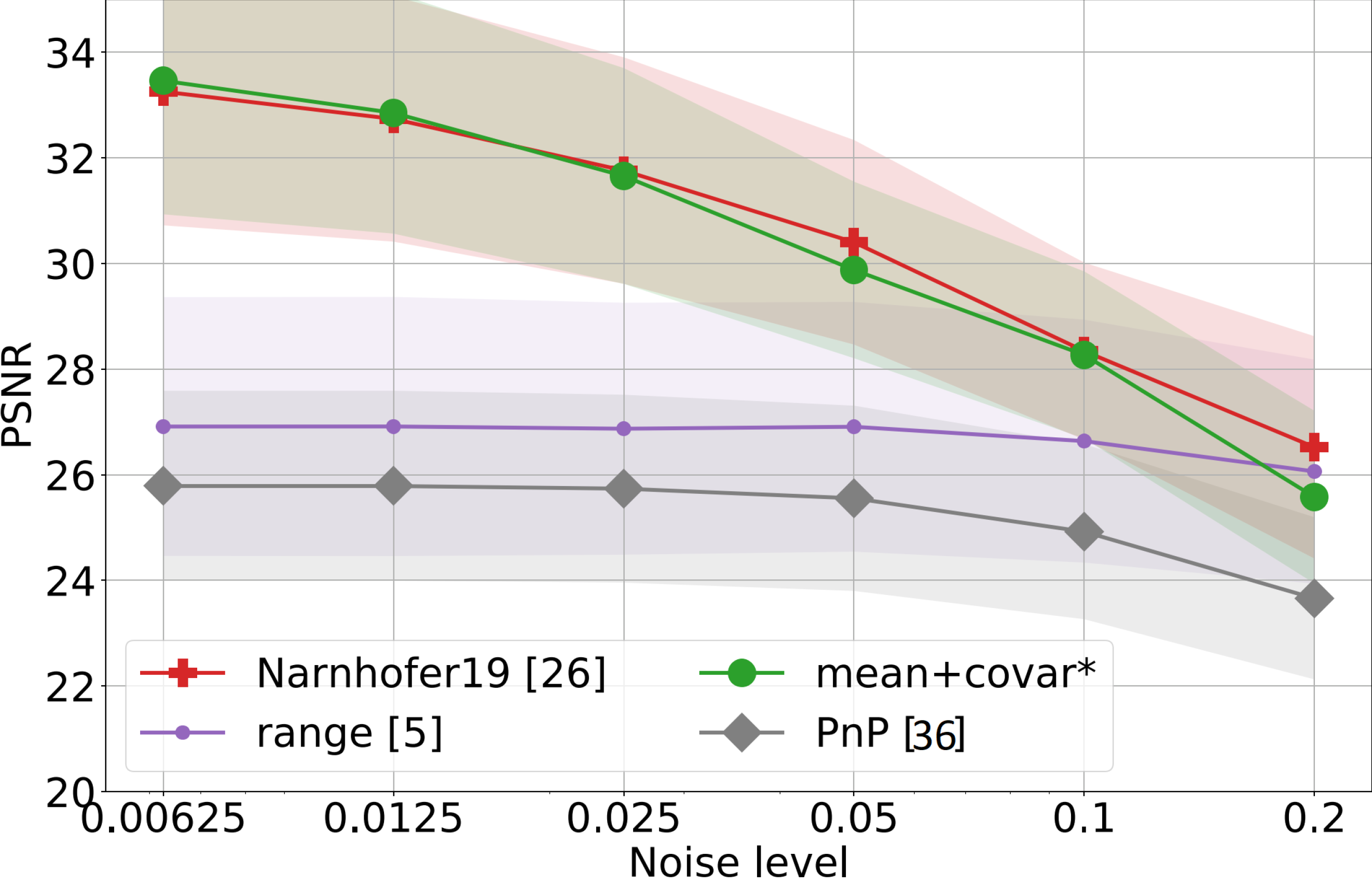}
		\caption{Comparison of the \textit{mean+covar*} method with \textit{range}~\cite{Bora2017} and \textit{Narnhofer19}~\cite{Narnhofer2019}, generative regularizer approaches. \changes{We also compare with the  \textit{PNP-ADMM method}~\cite{ryu2019plug}}. The experimental setup is as figure~\ref{fig:comparepsnr-unlearned}.}
		\label{fig:comparepsnr-priors}
	\end{figure*}

	Figure~\ref{fig:comparepsnr-unlearned} shows comparisons with TV and least squares reconstructions for a range of noise levels and the number of radial spokes in the sampling pattern. Image examples can be seen in columns 2 and 3 of figures~\ref{fig:Images example} and~\ref{fig:Images example noise}. We see the results of \textit{mean+covar*} track TV but with improvements across the range of radial spokes and noise levels tested. Especially in the examples in figure~\ref{fig:Images example noise} you can see the piece-wise constant shapes typical of a TV reconstruction, whereas for the same measured data  \textit{mean+covar*} has managed to reconstruct more of the fine detail.   As expected, due to the difficult nature of the inverse problem, the least squares reconstruction does poorly. As a rough indicator, our un-optimized implementation took 1.8 seconds for reconstructing one image using least squares, and 6.4 seconds using TV, for one choice of regularization parameter. For one choice of the regularization parameter, \textit{mean+covar*} took  78.2 seconds. 
	\paragraph{Comparison with other data-driven priors} 
	Comparisons with \textit{Narnhofer19}, \changes{ \textit{PnP-ADMM} }and \textit{range} are given in figure~\ref{fig:comparepsnr-priors} and columns 4 and 5 of figures~\ref{fig:Images example} and~\ref{fig:Images example noise}. We see that the results of searching in the \textit{range} of the generator saturate so that even with increased data in the form of radial spokes, or better data in the form of less noise, the reconstructions do not see significant improvement. The example image reconstructions reflect this; they show overly smoothed images that, although have the right structure, do not contain any of the fine detail of the ground truth. This could be because the generative model is not expressive enough to match this fine detail. Alternatively, it could be because the non-convex optimization has failed to find a good enough minimum to match the measured data. The results of \textit{Narnhofer19} are much more detailed and very similar in PSNR values to our \textit{mean+covar*}. In the images, there is some evidence of smoothing of details~\eg in figure~\ref{fig:Images example noise}. \changes{The \textit{PnP-ADMM} method did better for a higher number of radial spokes when we would expect the adjoint to be similar to images in the training set of high-quality reconstructions, however, it is still not competitive with our proposed method or  \textit{Narnhofer19}}. Our un-optimized implementation took 106.7 seconds for \textit{Narnhofer19}  and 23.9 seconds for the  \textit{range} method. \changes{The \textit{PnP-ADMM} method took 0.025 seconds to run, terminating at a small number of iterations.}

	\paragraph{Comparison with supervised end-to-end methods} 
	
\begin{figure*}
		\centering
		\includegraphics[width=0.45\linewidth]{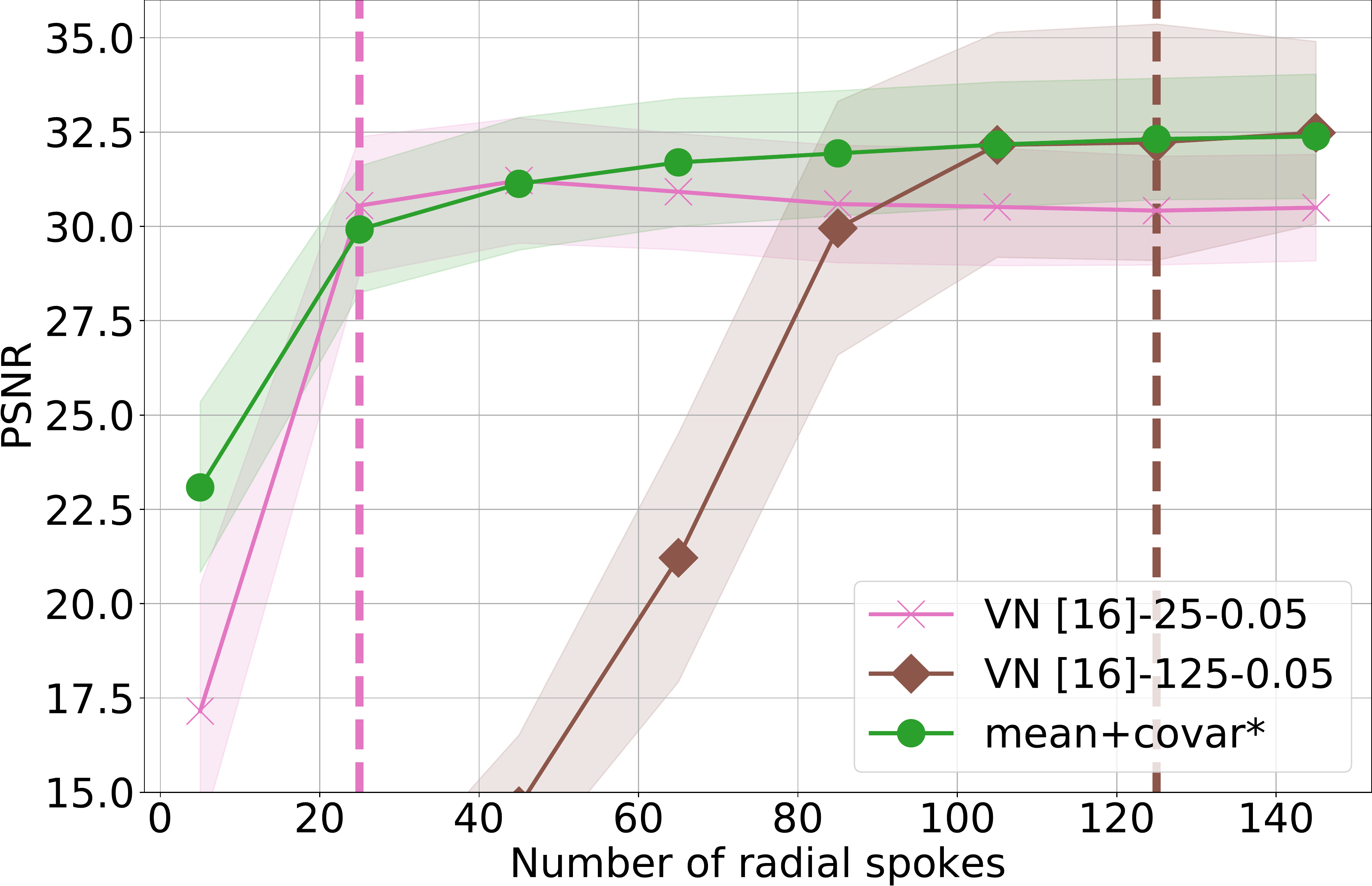}
		\includegraphics[width=0.45\linewidth]{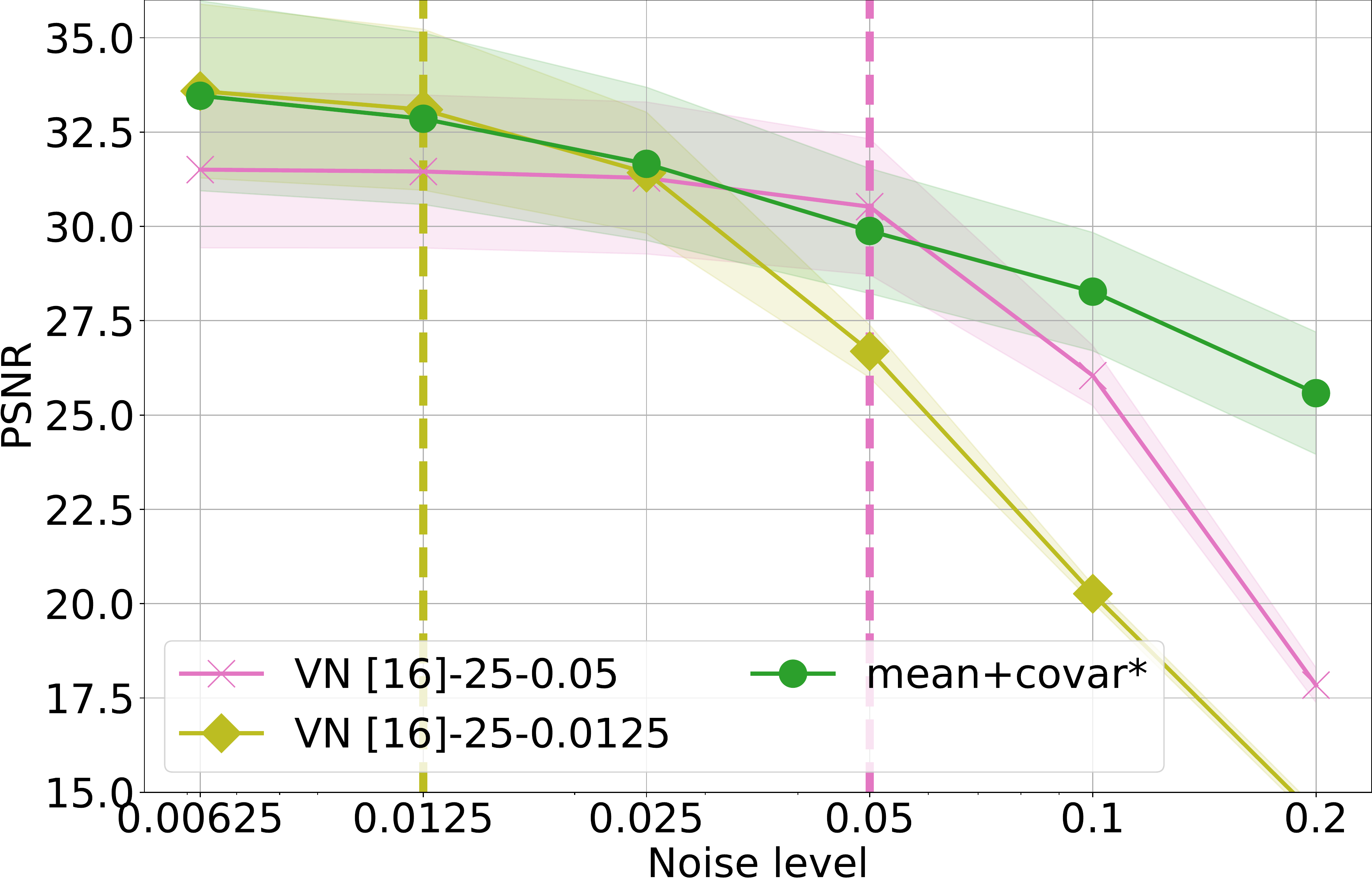}
		\caption{Comparison of the unsupervised \textit{mean+covar*} method with the supervised variational networks~\cite{Hammernik2018}. The experimental setup is as figure~\ref{fig:comparepsnr-unlearned} but in addition, the vertical lines depict the experimental settings the variational networks were trained on.}
		\label{fig:comparepsnr-end-to-end}
	\end{figure*}

	We now compare \textit{mean+covar*} with variational networks~\cite{Hammernik2018}, trained end-to-end\changes{, using paired training data obtained using} particular noise and sampling settings.  Figure~\ref{fig:comparepsnr-end-to-end} shows the PSNR results for 50 test images, with vertical lines highlighting the settings where the networks were trained. Example images are shown in columns 6 and 7 of figures~\ref{fig:Images example} and~\ref{fig:Images example noise} with the trained for setting highlighted in red.  We note that the variational networks achieve the best results, in comparison to the other methods, for the settings they were trained on.  This is as expected for end-to-end reconstructions. We see that the results are less optimal the further from the trained for setting, with some particularly poor results,~\eg in figure~\ref{fig:Images example}, while our unsupervised method remains consistent.  Also as expected, the variational networks are quick to implement once trained, taking 3.4 seconds for one image\changes{.}

\begin{figure*}
		\centering
		\includegraphics[width=\linewidth]{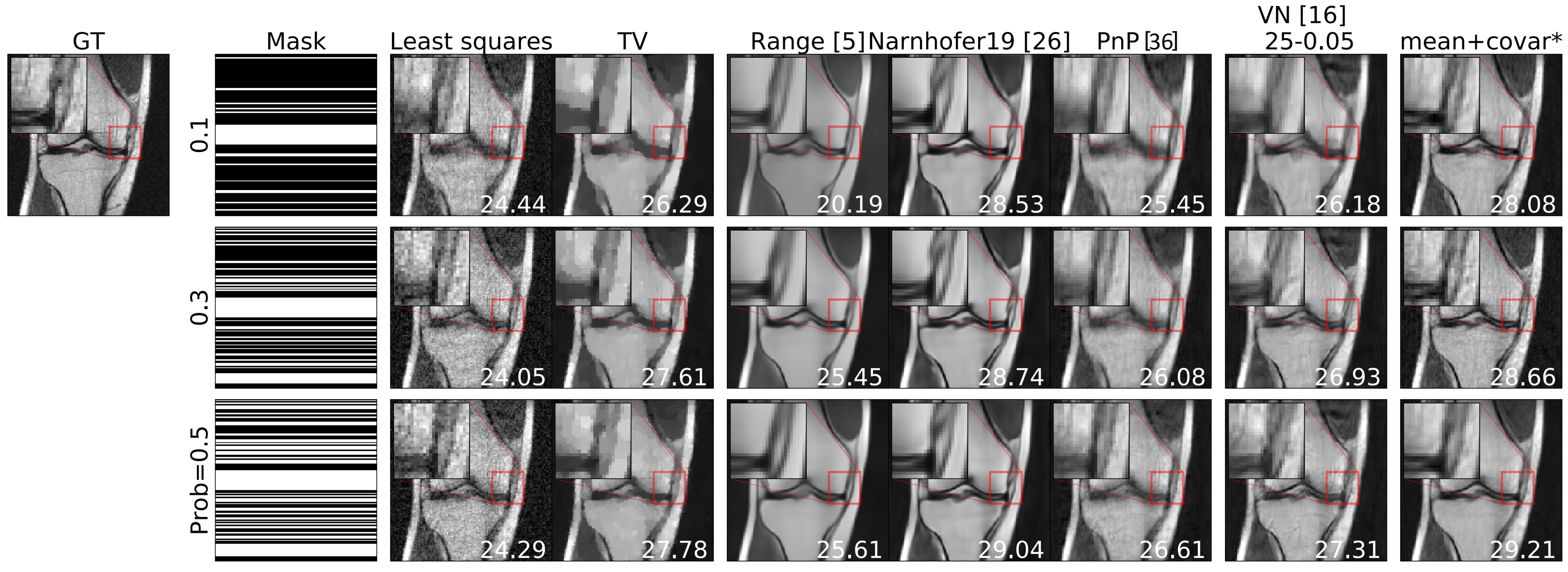}
		\caption{Example reconstructions of an image using measured data taken with different horizontal random sampling unseen in the training of the variational networks. The first column gives the k-space mask used to sample the data. The remaining columns give reconstructions for the different methods. }
		\label{fig:Images example hori}
	\end{figure*}

	\paragraph{Generalization to a Different Sampling Pattern}
Finally, we test the generalization ability of the methods by changing the sampling pattern. We mask horizontal rows in the k-space, taking 16 (out of 128) center rows and a uniformly random selection of other rows with a given probability, $p$. Fully sampled images correspond to $p=1$. Example sampling patterns and one example reconstruction is given in figure~\ref{fig:Images example hori}. We see an improvement over TV  for \textit{mean+covar*}. \textit{Narnhofer19} gives good PSNR values but the images are potentially over-smoothed. For example, the vertical lines in the top left part of the zoomed-in region are consistently missing from the \textit{Narnhofer19} reconstructions. The variational network has not been trained on this horizontal sampling pattern and although gives a reasonable reconstruction, seems to have some artifacts, for example in the top right and bottom left of the image.

	\section{Discussion \label{discussion}}
\noindent The numerical results show that the generative regularizer \textit{mean+covar*} consistently outperforms TV, \changes{as shown in} figure~\ref{fig:comparepsnr-unlearned}, and is competitive with other unsupervised generative model regularization schemes, figure~\ref{fig:comparepsnr-priors},    over a range of noise levels and sampling patterns.  This is an important contribution, demonstrating a generative model, trained on an MRI dataset, provides an effective prior for image reconstruction \changes{with varying levels of information}. As part of the ablation study, figures~\ref{fig:plot_generative_models} and~\ref{fig:ablation}, we demonstrated that models \textit{mean+diag} and \textit{mean+identity} provided some regularization but \textit{mean+covar*} provided the most flexibility to fit the  data and thus the best results. Searching in the \textit{range} of the generator we saw evidence that the generator was not expressive enough to fit the data and that the reconstructed images did not continue to improve with more or higher-quality data, this matches with similar results in~\cite{Bora2017, Duff2021}.

	The generative regularizers presented are not end-to-end methods and still require the use of an iterative optimization scheme to reconstruct an image given an observation, furthermore the optimization objective is non-convex.  \changes{Fortunately,} this optimization is practically straightforward as the gradients can be calculated automatically,~\eg using TensorFlow, and the optimization can be initialized with the adjoint reconstruction. The benefit of the unsupervised approach is that retraining is not required for different forward models. Figure~\ref{fig:comparepsnr-end-to-end} shows that the variational network end-to-end method~\cite{Hammernik2018}, although provides the best results on the forward model it was trained on,  sees its success drop off as you test it on other forward model settings. An additional benefit of \textit{generative regularizers} is that you can visually inspect the learned image prior and we demonstrate this in figure~\ref{fig:learned-covar}.

Training the Cholesky weights for the precision matrix proved tricky.  As shown in figure~\ref{fig:plot_generative_models}, the structured covariance model has learned some of the residual structure that we expect but at the expense of applying higher variances across the whole image. This may make the structured covariance model too permissive, allowing it to fit noise or artifacts from the measurements. With an improved model, we expect that the regularization parameters in~\eref{reg2} would not be required, they could be set by the hierarchical Bayesian derivation.  Future work could consider what additional priors or alternative modeling schemes could help the learning of this structured covariance model, such as described in~\cite{Dorta2018a}.

		{Although this work demonstrates a data-driven  approach to inverse problems that is flexible and powerful there are several ways the MRI simulation could be made more realistic}. {As discussed in section~\ref{experiments}, our experiments are inspired by the single coil fastMRI challenge which uses \changes{magnitude} images and, as the name suggests, a single-coil acquisition. In reality, modern MRI scans use a multi-coil acquisition and usually produce complex images with a non-trivial phase. A multi-coil acquisition could be incorporated into our framework with a change of the forward model,~\eg via a SENSE formulation~\cite{Pruessmann1999}.   Modeling the structures and correlations between the real and complex parts of an MR image is an interesting open problem and subject to future work. 
		Finally, the radial masks we used to simulate a radial sampling pattern, were just an approximation of the actual MR physics and future work could consider bringing this closer to real-world applications. }
		
		{As in most machine learning approaches, we} have assumed that the images we wish to reconstruct are similar to those in the training dataset used to train the generative model. This is not obvious in medical imaging, where damage, tumors, and illness can lead to different image presentations that may be far from those of healthy volunteers used to create datasets. Modeling this,~\eg by sparse deviations~\cite{Dhar2018, Duff2021} away from the range of a generative model,  is an interesting area of future development. 
  
\changes{Another consideration for future work is to consider going beyond the MAP estimate and considering quantifying uncertainty in the reconstructed image. One possible avenue would be to produce a range of reconstructions by sampling from the learned decoder for an optimized value of $z$. Further work would be required to interpret and present these results in a meaningful and theoretically justified way.  }
	\section{Conclusion} \noindent 	Generative regularizers provide a bridge between black-box deep learning methods and traditional variational regularization techniques for image reconstruction. We propose \changes{an} adaptive generative regularizer that models image correlations with a structured noise network.    The proposed method is trained unsupervised, using only high-quality reconstructions and is thus adaptable to different noise and k-space sampling levels.   Our results show that generative regularizers are most effective when the underlying generative model outputs both an image but also a non-trivial covariance matrix for each point in the latent space. The covariance provides a learned metric that guides where the reconstruction can or cannot vary from the learned generative model. We demonstrated the success of this approach through comparisons with other unsupervised and supervised approaches. 
 
\section*{Acknowledgments}

 MD is supported by a scholarship from the EPSRC Centre for Doctoral Training in Statistical Applied Mathematics at Bath (SAMBa), under the project EP/L015684/1. MJE acknowledges support from the EPSRC (EP/S026045/1, EP/T026693/1, EP/V026259/1) and the Leverhulme Trust (ECF-2019-478).
NDFC acknowledges support from the EPSRC CAMERA Research Centre (EP/M023281/1 and EP/T022523/1) and the Royal Society.

\printbibliography

\end{document}